\documentstyle[12pt]{article}

 \newcommand{\be}{\begin{equation}}
 \newcommand{\ee}{\end{equation}}
 \newcommand{\ba}{\begin{eqnarray}}
 \newcommand{\ea}{\end{eqnarray}}
 
 \newcommand{\del}{\partial}

\newcommand{\lef}{\left}

\newcommand{\ri}{\right}

\newcommand{\cl}{{\cal L}}

\newcommand{\fr}{\frac}

\begin{document}

\begin{titlepage}

\topmargin -15mm

\vskip 10mm

\centerline{ \LARGE\bf Creation Operators for 2-branes and Duality}

\vskip 2mm

\centerline{ \LARGE\bf in BF and Chern-Simons Theories in D=5}

    \vskip 2.0cm

    \centerline{\sc E.C.Marino }

     \vskip 0.6cm
     
    \centerline{\it Instituto de F\'\i sica }
    \centerline{\it Universidade Federal do Rio de Janeiro }
    \centerline{\it Cx.P. 68528}
    \centerline{\it Rio de Janeiro RJ 21941-972}
    \centerline{\it Brazil}


\vskip 1.0cm

\begin{abstract} 
 
We explicitly construct the creation operators for the quantum field configurations 
associated to quantum membranes (2-branes) in BF and generalized Chern-Simons theories
in a spacetime of dimension D=5. The creation operators for quantum excitations carrying 
topological charge are also obtained in the same theories. For the case of 
D=5 generalized Chern-Simons theory, we show that this operator actually creates an
open string with a topological charge at its tip. It is shown that a duality structure 
exists in general, relating the membrane and topological excitation operators and the 
corresponding dual algebra is derived. Composite topologically charged membranes
are shown to possess generalized statistics that may, in particular, be fermionic.
This is the first step for the bosonization procedure in these theories. Potential 
applications in the full quantization of 2-branes is also briefly discussed.
			      
\end{abstract}

\end{titlepage}

\hoffset= -10mm

\leftmargin 23mm

\topmargin -8mm
\hsize 153mm
 
\baselineskip 7mm
\setcounter{page}{2}

\section{Introduction}
\setcounter{equation}{0}

Extended objects such as strings and membranes have been playing an 
important role in recent theoretical advances in high-energy physics \cite{pol}
and cosmology \cite{sv}. When coupled
to field theories, it happens that specific field 
configurations become attached to these objects, sometimes
substantially influencing their physical properties. These field configurations, 
frequently carry a nontrivial topological charge, thereby 
effectively transforming the extended object into a topological excitation.
Such excitations usually present the interesting feature of acting as
disorder variables, being dual to their counterparts, the order variables, which
can be expressed in terms of the lagrangian fields of the theory. 
This duality relation, physically,
means that the vacuum expectation values of the corresponding operators
may be used to characterize
the different possible phases of the system, 
according to the breakdown of the corresponding symmetries.
It is easy to understand the importance of the role played by topological excitations in this
framework by recalling that the presence of these in the physical spectrum of excitations
is closely related to the occurence of a spontaneously broken symmetry. Mathematically,
duality is expressed as an algebraic relation between the creation operators associated
to the order and disorder variables, the so called dual algebra, which has far reaching
effects in the physical properties of the system. One possibility opened by the
existence of a dual structure manifested through such an algebraic relation is the
method of bosonization, by which fermions are mapped into bosons and vice-versa. 

As a consequence of the presence of field configurations that become intrinsically
attached to an extended object, its statistics may change from
bosonic to a generalized one and, in particular, they may become fermionic.
This is the statistical transmutation, first obtained for point particles in D=3 \cite{st}
an subsequently generalized for strings in D=4 \cite{stst,em1}.
Bosonization, on the other hand, is a different 
but related procedure, by which we combine order and
disorder dual variables in order to produce a new excitation with generalized 
statistics that may, in particular be fermionic \cite{mand,emsw}.
Indeed, the construction of a fermionic operator associated
to a certain object, completely expressed in terms of the bosonic fields of the 
underlying field theory to which this object is coupled is in fact the first step
towards bosonization. Subsequently, the fermionic correlation functions should be
reproduced in the framework of the bosonic theory. The bosonization
program has been completely implemented in a spacetime of dimension D=2 \cite{mand,bos},
where it has allowed, for instance, the obtainment of exact quantum operator solutions of
nonlinear theories. In D=3 it has been only partially fulfiled \cite{embos,bos3}. 

The formulation of the above mentioned ideas both for objects of a 
dimension larger than one (strings) and spacetimes higher than D=4 is a natural extension
of this line of investigation that may have very interesting and useful 
consequences. In the present work, after presenting an unified approach to the subject
in dimensions D=2,3 and 4, we investigate the case of membranes and strings in a 
spacetime of dimension D=5, considering specifically BF and generalized Chern-Simons
theories. We explicitly construct, in both cases, the creation operators for 
quantum excitations bearing a nonzero topological charge. In the specific case
of the generalized Chern-Simons theory, we also consider the case where
this excitation is an open string with a topological charge at its tip. We also
explicitly obtain the creation operators of the field configurations associated
to quantum membranes and show that indeed they are the eigenstates of the field
operators with the correct eigenfield configurations. In both theories, we
study the duality relation and derive the dual algebra satisfied by the creation 
operators of topologically charged states and the quantum membrane field configurations.
Based on these dual algebras, we then proceed to the obtainment of composite
topologically charged membranes and show that they possess general statistics that may
be, in particular, fermionic. In this way, therefore, the first step in the 
bosonization program may be accomplished in the theories considered here.
Perspectives for future extensions os this work are presented in Section 5.

\section{Overview of Quantization of Topological Excitations, p-branes and Duality}
\setcounter{equation}{0}

\subsection{Topological Charges, Magnetic Fluxes and p-branes}

The identically conserved topological current is expressed in a D-dimensional
spacetime in terms of a tensor field of rank D-2. In D=2,3,4 and 5, it is given,
respectively by
$$
J^\mu = \epsilon^{\mu\nu}\del_\nu \phi
$$
$$
J^\mu = \epsilon^{\mu\alpha\beta}\del_\alpha B_\beta
$$
$$
J^\mu = \epsilon^{\mu\nu\alpha\beta}\del_\nu B_{\alpha \beta}
$$
\be
J^\mu = \epsilon^{\mu\lambda\nu\alpha\beta}\del_\lambda B_{\nu\alpha \beta}
\label{1}
\ee
The generalization to higher dimensions is obvious. The topological charge, in 
all cases is given by
\be
Q = \int d^{D-1}x J^0
\label{2}
\ee
The magnetic field, or magnetic flux density, on the other hand, is always given in 
terms of the spatial components of the vector field $A_\mu$. Indeed, we have
$$
{\cal B} = \epsilon^{ij} \del_i A_j
$$
$$
{\cal B}^i = \epsilon^{ijk} \del_j A_k
$$
\be
{\cal B}^{ij} = \epsilon^{ijkl} \del_k A_l ,
\label{3}
\ee
respectively, in three, four and five spacetime dimensions. Observe that only in D=4
the magnetic field is a vector, whereas in D=2 it is not defined.

Let us consider now the current density associated to general extended objects (p-branes), 
namely strings, membranes or even particles, in any
spacetime dimension D. For a particle (0-brane) 
in the point $x_0$, the current density is given by
\be
j^\mu (x;x_0) = \int_{L(x_0)} d \xi^\mu \delta^D (x-\xi) ,
\label{4}
\ee
where $L(x_0)$ is the universe-line of the particle. For a string (1-brane) 
along the line $L$, we have
\be
j^{\mu\nu} (x;L) = \int_{S(L)} d^2 \xi^{\mu\nu} \delta^D (x-\xi) ,
\label{5}
\ee
where $S(L)$ is the universe-sheet of the string at $L$. Finally, for a membrane
(2-brane) along the surface $S$, we have accordingly,
\be
j^{\mu\alpha\beta} (x;S) = \int_{V(S)} d^3 \xi^{\mu\alpha\beta} \delta^D (x-\xi) ,
\label{6}
\ee
where $V(S)$ is the universe-volume of the membrane at $S$.

In what follows, we are going to explore the interplay between the topological charge, 
magnetic flux and extended objects in BF and Chern-Simons theories in general. 
Observe that the particle, string and membrane densities associated
respectively to the currents (\ref{4}), (\ref{5}) and (\ref{6}) are a scalar, a vector
and a rank-2 tensor, respectively $j^0$, $j^{0i}$ and $j^{0ij}$. These is precisely
the nature of the magnetic field in D=2, 3 and 4 respectively, according to (\ref{3}).
This is not a mere coincidence. It turns out that when particles, strings or membranes are
coupled to generalized BF or Chern-Simons theories in D dimensions, a magnetic field
(tensor in general) is imparted to the respective object.

It happens also, that a duality relation exists at the quantum level, 
between the magnetic field bearing object and the topological charge bearing states
corresponding to (\ref{1})that can be constructed in the above mentioned theories.

\subsection{Sequences of BF and Chern-Simons Theories}

BF theories can be constructed in any spacetime dimension D, out of a vector field
$A_\mu$ and a (D-2)-rank tensor field $B_{\mu_1...\mu_{D-2}}$. In D=2,3, 4 and 5, respectively,
we have 
\be
\cl_2 = - \fr{1}{4} F_{\mu\nu}^2 - \epsilon^{\mu\nu} 
A_\mu \del_\nu \phi + \fr{1}{2} \del_\mu \phi \del^\mu \phi,
\label{7a}
\ee
\be
\cl_3 = - \fr{1}{4} F_{\mu\nu}^2 -  \epsilon^{\mu\alpha\beta} 
A_\mu \del_\alpha B_\beta - \fr{1}{4} H_{\mu\nu}^2,
\label{7}
\ee
\be
\cl_4 = - \fr{1}{4} F_{\mu\nu}^2 - \fr{1}{2} \epsilon^{\mu\nu\alpha\beta} 
A_\mu \del_\nu B_{\alpha\beta} + \fr{1}{12} H_{\mu\alpha\beta}^2
\label{8}
\ee
and
\be
\cl_5 = - \fr{1}{4} F_{\mu\nu}^2 - \fr{1}{2} \epsilon^{\mu\lambda\nu\alpha\beta} 
A_\mu \del_\lambda B_{\nu\alpha\beta} + \fr{1}{24} H_{\mu\nu\alpha\beta}^2
\label{9}
\ee
where
$H_{\mu\nu} = \del_\mu B_{\nu} - \del_\nu B_{\mu}$, 
$H_{\mu\alpha\beta} = \del_\mu B_{\alpha\beta} + {\rm cyc.\  perm.\ }  (\mu\alpha\beta)$
$H_{\mu\nu\alpha\beta} = \del_\mu B_{\nu\alpha\beta} + {\rm cyc.\  perm.\ }  
(\mu\nu\alpha\beta)$ are the field intensity tensors of the B-field in
D=3,4 and 5, respectively and $F_{\mu\nu} = \del_\mu A_{\nu} - \del_\nu A_{\mu}$. A common 
feature of thte BF theories in any dimension is that integration over each of the fields
$A$ or $B$ produces a mass term for the other.

The field equations obtained by varying with respect to $A_\mu$
in the three lagrangians above are
\be
\del_\nu F^{\mu\nu} = J^\mu ,
\label{10}
\ee
where $J^\mu$, the topological current, is given by each of the expressions in
(\ref{1}), respectively in D=2,3, 4 and 5.
We immediately see that the topological charge is the source of the electromagnetic field
in BF theories in any dimension. Varying with respect to $B$, in (\ref{7}), (\ref{8})
and (\ref{9}), on the other hand, we get
\be
\del_\alpha H^{\alpha\mu} = \epsilon^{\mu\nu\beta} \del_\nu A_\beta ,
\label{11a}
\ee
\be
\del_\alpha H^{\alpha\mu\nu} = \epsilon^{\mu\nu\alpha\beta} \del_\alpha A_\beta ,
\label{11}
\ee
\be
\del_\alpha H^{\alpha\mu\nu\lambda} = 
\epsilon^{\mu\nu\lambda\alpha\beta} \del_\alpha A_\beta .
\label{12}
\ee
It is clear that the magnetic fields given by 
(\ref{3}), in D=3,4 and 5, are the source of the ``electric'' B-field.
The case of three spacetime dimensions is special because ${\cal B} = J^0$ and therefore
magnetic flux is identical to topological charge. This fact is responsible for many 
of the interesting features of 3-dimensional Chern-Simons theory as we shall see below.          

Generalized Chern-Simons theories may also be constructed in a spacetime of arbitrary 
dimension D, out of the same fields used for the construction of the BF theories 
above. Including the minimal coupling to particle, string and membrane densities, we
have, respectively, in D=3, 4 and 5,
\be
\cl_{CS3} = \fr{1}{2} \epsilon^{\mu\alpha\beta} 
A_\mu \del_\alpha A_\beta - j^\mu A_\mu ,
\label{13}
\ee
\be
\cl_{CS4} =  \epsilon^{\mu\nu\alpha\beta} 
A_\mu \del_\nu B_{\alpha\beta} - j^{\mu\nu} B_{\mu\nu} - j^\mu A_\mu 
\label{14}
\ee
and
\be
\cl_{CS5} =  \epsilon^{\mu\lambda\nu\alpha\beta} 
A_\mu \del_\lambda B_{\nu\alpha\beta} - j^{\nu\alpha\beta} B_{\nu\alpha\beta}- j^\mu A_\mu 
\label{15} .
\ee
Notice that the case D=3 is special because, both $A_\mu$ and $B_\mu$ being vector fields,
there are three possible equivalent topological terms involving $AA$, $AB$ and $BB$, 
respectively. Making linear combinations of these two fields, however, it is possible to write
all of them as a unique topological term as in (\ref{13}).

Varying each of the lagrangians above with respect to $A_\mu$, we obtain
\be
j^\mu = J^\mu ,
\label{16}
\ee
where $J^\mu$ is the topological current, in each case, given by Eqs. (\ref{1}). We see
that the topological charge density becomes identified with the particle density
coupled to the Chern-Simons theory. In the specific case of D=3, as is well known, the 
topological charge density is also the magnetic flux density or magnetic field (a scalar
in this case). This fact is responsible for the statistical transmutation of particles
coupled to the Chern-Simons field in three dimensions \cite{st}.

If we vary (\ref{14}) and (\ref{15}) with respect to the B-field, on the other hand, we 
obtain 
\be
j^{\mu\nu} =   \epsilon^{\mu\nu\alpha\beta} \del_\alpha A_\beta
\label{17}
\ee
and
\be
j^{\mu\nu\lambda} =   \epsilon^{\mu\nu\lambda\alpha\beta} \del_\alpha A_\beta
\label{18}
\ee
Taking the 0-component of each of the currents above, that corresponds
to the string and membrane densities, respectively, we get
$$
j^{0i} =  {\cal B}^i 
$$
\be
j^{0ij} = {\cal B}^{ij} 
\label{19}
\ee
These expressions clearly show that a magnetic field is associated to the p-brane 
(string or membrane) coupled to the B-field in a generalized Chern-Simons 
theory.

In the next subsection, we show how to construct the quantum field operators that create
states bearing topological charge and also the operators creating the quantum field 
configuration that are eingenstates of the magnetic field operator with the 
eigenfield configurations corresponding to the ones associated to strings and membranes 
in the theories above. We also investigate the duality relation existing among them.

\subsection{Creation Operators of Topological Excitations and Extended Objects }

Let us examine here the general form of the topological excitation and extended
object creation operators
both in Chern-Simons and BF theories in arbitrary dimensions. We start with 
the creation operators of
topological charge bearing excitations in generalized BF-theories.
The best known case is D=2, where this operator is given by \cite{mand}
\be
\mu(x,t) = \exp \lef \{-i a \int_{-\infty}^x d \xi \Pi (\xi, t) \ri \}
\label{20}
\ee
where $\Pi = \del_0 \phi$ is the momentum canonically conjugate to $\phi$.
This expression has been generalized to three and four spacetime dimensions,
respectively, in the form
\be
\mu(\vec x,t) = \exp \lef \{-i a \int_{-\infty}^{\vec x} d \xi^i 
\epsilon^{ij} \Pi^j (\vec \xi, t) \ri \}
\label{21}
\ee
and
\be
\mu(\vec x,t) = \exp \lef \{-i a \int_{-\infty}^{\vec x} d \xi^i 
\epsilon^{ijk} \Pi^{jk} (\vec \xi, t) \ri \}, 
\label{22}
\ee
where $\Pi^j = H^{oj}$ and $\Pi^{jk} = H^{ojk}$. $H^{\mu\nu}$ is the field intensity 
tensor of the $B_\mu$ field in D=3 and $H^{\mu\nu\alpha}$ the one of the 
$B_{\mu\nu}$ field in D=4. All of these three 
operators create eigenstates of Q, Eq. (\ref{2}),
with eigenvalue proportional to $a$ (see \cite{em1}, for D=4 and \cite{em2}, for D=3). 
In all cases, D=2,3 and 4 above, the operator $\mu$ can be written as 
\be
\mu(x) = \exp \lef \{-i a \int_{-\infty}^x dx^\mu J_\mu \ri \},
\label{23}
\ee
with $J^\mu$ given by the corresponding expression in (\ref{1}).

Let us consider now the topological charge creation operators in generalized Chern-Simons
theories given by (\ref{13})-(\ref{14}).
According to the results of \cite{em4,bm} the operator that creates eigenstates of the
topological charge $Q$, Eq. (\ref{2}) is given by the expression
\be
\mu(\vec x,t) = \exp \lef \{-i a \int_{-\infty}^{\vec x} d \xi^\mu A_\mu (\vec \xi, t) \ri \}
\label{24}
\ee
in both theories, respectively in D=3 and D=4.
In the following sections, we are going to determine the creation operator of
topological charge eigenstates, both in BF and Chern-Simons theories in D=5.

We have seen above that a magnetic field, which may be a 
scalar, vector or tensor, given by (\ref{3}),
is associated to a particle, string or membrane in generalized Chern-Simons 
theories in D=3, 4 and 5, respectively. The operator creating the quantum magnetic 
field eigenstates associated to these objects has been studied in
\cite{em4} for the case D=3 and in \cite{bm} for D=4. For Chern-Simons theory in
D=3, this operator is given by \cite{em4}
\be
\sigma(x) = \exp \lef \{-i b \int d^3 x  j^\mu A_\mu  \ri \} ,
\label{25}
\ee
where the current $j^\mu$ is given by (\ref{4}). It can be shown \cite{em4}
that 
\be
{\cal B} |\sigma > = b |\sigma >
\label{26}
\ee
Inserting (\ref{4}) in (\ref{25}), we immediately find that in this case
$\sigma(x) = \mu(x)$, reflecting, at the quantum level,  
the fact that (scalar) magnetic field and topological 
charge are identified in D=3 Chern-Simons theory.

For Chern-Simons theory in D=4, conversely, the operator that creates
the eigenstates of the quantum vector magnetic field that is associated to
a string coupled to the tensor field according to (\ref{14}), is given by
\cite{bm}
\be
\sigma(L) = \exp \lef \{-i b \int d^4 x  j^{\mu\nu}(L) B_{\mu\nu}  \ri \} ,
\label{27}
\ee
for a string at the curve $L$, associated to the density $j^{\mu\nu}$ given by
(\ref{5}). It can be shown \cite{bm} that for a closed spatial string at the curve $C$
we have, indeed,
\be
{\cal B}^i |\sigma (C) > = 2 b 
\lef [ \oint_C d \xi^i  \delta (\vec \xi - \vec x) \ri  ] |\sigma (C) > . 
\label{28}
\ee

Interestingly, in the case of BF theories in D=3 and D=4, the creation operator of the
extended objects carrying the corresponding attached magnetic field has exactly the 
same form as the ones in (\ref{25}) and (\ref{27}), respectively in D=3, \cite{em3}
and D=4, \cite{em1,em5}. In this case they create eigenstates of the generalized
``electric'' B-field, which is precisely the magnetic field of $A_\mu$,
as we have seen above. 
Of course, in D=3  $A_\mu$ and $B_\mu$ are identified. 

These observations allow us to induce that the general
form of the operator creating the quantum field states associated to general extended
objects, either in Chern-Simons or BF-theories, in an arbitrary dimension, 
should be given by
\be
\sigma = \exp \lef \{-i b \int d^D x  j^{\mu ...\nu} B_{\mu ...\nu}  \ri \} ,
\label{29}
\ee
where the object density $ j^{\mu ...\nu}$ is given by (\ref{4}), (\ref{5}), (\ref{6}) 
anf generalizations thereof. We are going to verify this explicitly fore case
of membranes in D=5, in the following two sections.

\subsection{Duality}

There exists a remarmakable relationship between the creation operators of topological 
charge and extended objects in a given theory. This is order-disorder duality,
in the sense of Kadanoff and Ceva \cite{kc}. Physically, the 
vacuum expectation value of the topological charge
creation operator behaves as a disorder parameter, dual to the corresponding
expectation value of the extended object creation operator. Observe that
in the case of a vector field (\ref{29}) is nothing but the Wilson loop (or line)
operator. This duality has been extended and explored in continuum field theory by
't Hooft \cite{th}.

From the mathematical point of view, order-disorder duality is expressed as an
algebraic relation beteen $\sigma$ and $\mu$, as first discovered 
in the framework of the Ising model \cite{kc},
and subsequently generalized to field theory \cite{th}.
In D=2, the dual algebra is \cite{emsw,da1d}
\be
\mu(x,t) \sigma(y,t) = \exp \lef \{ i\fr{ab}{2 \pi} \theta(y-x)\ri \} \sigma(y,t) \mu(x,t) ,
\label{30}
\ee
where $\theta$ is the Heaviside function.
In D=3, the corresponding algebraic relation is \cite{em3,em4}
\be
\mu(\vec x,t) \sigma(\vec y,t) = \exp \lef \{ i\fr{ab}{2 \pi} {\rm arg}(\vec y- \vec x)
\ri \} \sigma(\vec y,t) \mu(\vec x,t) .
\label{31}
\ee
The generalization of these algebraic relations existing between 
a point topological charge operator and the creation operator of a 
string at a spatial curve $C$ for D=4 has been obtained in \cite{em1,bm}
and is given by
\be
\mu(\vec x,t) \sigma(C,t) = \exp \lef \{ i\fr{ab}{4 \pi} \Omega (\vec y;C) 
\ri \} \sigma(C,t) \mu(\vec x,t) ,
\label{32}
\ee
where $\Omega (\vec y;C)$ is the solid angle comprised by $\vec y$ and the spatial 
curve $C$.

One of the most important consequences of the dual (order-disorder) algebra is the 
possibility of constructing a mapping between fermionic and bosonic fields
known as bosonization \cite{mand,bos}. Indeed, it follows from the above relations
that the product of $\sigma $ and $\mu$ fields may have 
generalized and, in particular, fermionic statistics. 
This product is precisely what appears in the
bosonized expression of the Dirac field in D=2, obtained by Mandelstam \cite{mand}.
Also in D=3, a similar product has been obtained in the bosonization of the
free massless Dirac field \cite{embos}.

\section{Quantum Magnetic Membranes and Topological Excitations in the BF Theory in D=5}
\setcounter{equation}{0}

\subsection{The Topological Charge Operator}

Let us investigate here the creation operator for quantum topological charge
eigenstates in the five dimensional BF theory described by the lagrangian (\ref{9}). 
Following the sequence expressed in (\ref{20}), 
(\ref{21}) and (\ref{22}), we write
\be
\mu(\vec x,t) = \exp \lef \{-i \fr{a}{3\!} \int_{-\infty}^{\vec x} d \xi^i 
\epsilon^{ijkl} \Pi^{jkl} (\vec \xi, t) \ri \}, 
\label{33}
\ee
where $\Pi^{jkl} = H^{jkl0}$ is the momentum canonically conjugate to $B^{jkl}$, 
which satisfies the following canonical commutation rules
\be
[B^{ijk}(\vec x, t), \Pi^{rst}(\vec y, t)] = {\rm i} \Delta^{ijk,rst} 
\delta^4(\vec x -\vec y) ,
\label{34}
\ee
where 
\be
\Delta^{ijk,rst} = \Sigma_{{\rm permutations}\ [rst]} 
\ (-1)^P \delta^{ir} \delta^{js} \delta^{kt},
\label{35}
\ee
$P$ being the parity of the permutation. These have been obtained by eliminating the 
second class constraints, without the use of any gauge condition. 

Let us determine the commutation of $\mu$ with the topological charge operator
\be
Q = \int d^{4}x \epsilon^{ijkl} \del_i B_{jkl}
\label{36}
\ee
Using (\ref{34}) and the Baker-Hausdorf formula, we readily get
\be
[Q,\mu] = a \mu ,
\label{37}
\ee
which implies that the states $|\mu> = \mu |0>$, created by $\mu$
are indeed eigenstates of the topological charge $Q$, given by (\ref{36}), 
with eigenvalue $a$.

\subsection{The Magnetic Membrane Creation Operator}

Let us study here the operator creating the quantum field configuration corresponding
to a magnetic membrane in D=5 BF theory. Following the sequence occurrying in
(\ref{25}), (\ref{27}) and (\ref{29}), we write, for a membrane at a spatial surface $S$,
\be
\sigma(S) = \exp \lef \{-i b \int d^5 x  j^{\mu\nu\alpha}(S) B_{\mu\nu\alpha}  \ri \} ,
\label{38}
\ee
where $j^{\mu\nu\alpha}(S)$ is given by (\ref{6}). Using (\ref{6}) and performing the 
$x$ integration in the above equation, we find
\be
\sigma(S;t) = \exp \lef \{- i b \int_{V(S)} d^3 \xi^{ijk}  B_{ijk} (\vec \xi, t)  \ri \} ,
\label{39}
\ee
where $V(S)$ is the universe-volume of the membrane at $S$, which has precisely $S$ as its
border and $d^3 \xi^{ijk}$, its volume element.

Let us determine the commutation relation between $\sigma(S;t)$ and the 
the 2-tensor magnetic field ${\cal B}^{ij}$, expressed by (\ref{3}), in D=5.
From (\ref{12}), it becomes clear that in D=5 BF theory, we have
\be
{\cal B}^{ij} = \del_k \Pi^{kij}
\label{40}
\ee
where $\Pi^{jkl} = H^{jkl0}$ is the momentum canonically conjugate to $B^{jkl}$. 

Let us call $\sigma(S;t) \equiv e^{\alpha (S;t)}$. Using (\ref{40}) and (\ref{34}),
we get
$$
[{\cal B}^{ij}(\vec x;t), \alpha(S;t)] = b \int_{V(S)} d^3 \xi^{ijk} \del_k 
\delta^4 (\vec \xi - \vec x)
$$
\be
= b \oint_S d^2 \xi^{ij} \delta^4 (\vec \xi - \vec x) ,
\label{41}
\ee
where, in the last step, we used the generalized Stokes' theorem. Notice that the
directions associated to $i,j$ are tangent to the membrane at each point.
Using the Baker-Hausdorf formula, we immediately find
\be
[{\cal B}^{ij}(\vec x;t), \sigma(S;t)] = 
b \lef [\oint_S d^2 \xi^{ij} \delta^4 (\vec \xi - \vec x)\ri ]  \sigma(S;t),
\label{42}
\ee
which implies
\be
{\cal B}^{ij}(\vec x;t) |\sigma(S;t)> = 
b \lef [\oint_S d^2 \xi^{ij} \delta^4 (\vec \xi - \vec x)\ri ] | \sigma(S;t)> .
\label{43}
\ee
This shows that $\sigma(S;t)$ creates eigenstates of the magnetic field with an
eigenconfiguration that is nonvanishing along the membrane. Integrating the
magnetic field along a surface $R$ orthogonal to the membrane, we get the magnetic
flux 
\be
\Phi_R = \int_R d^2\eta^{ij} {\cal B}^{ij}
\label{43a}
\ee
satisfying
\be
\Phi_R |\sigma(S;t)> = b |\sigma(S;t)> .
\label{44}
\ee

\subsection{Duality, Generalized Statistics and Fermionization}

Let us show now that a duality relation that generalizes (\ref{30}), (\ref{31}) and
(\ref{32}) can be established between the topological charge creation operator
$\mu$, given by (\ref{33}) and the membrane operator $\sigma$, given by (\ref{39}),
in the framework of BF theory in five spacetime dimensions.
Indeed, calling $\mu(\vec x;t) \equiv e^{\beta (\vec x;t)}$ and again 
$\sigma(S;t) \equiv e^{\alpha (S;t)}$, we obtain, using (\ref{34}),
\be
[\beta (\vec x;t), \alpha (S;t) ] = - {\rm i} ab  \int_{V(S)} d^3 \eta^i
\int_{-\infty}^{\vec x} d \xi^i \delta^4 (\vec \xi - \vec \eta) ,
\label{45}
\ee
where $d^3 \eta^i$ is the volume element of $V(S)$.
Using the result (\ref{57}) of the Appendix, we get
$$
[\beta (\vec x;t), \alpha (S;t) ] =  {\rm i} \fr{12 ab}{4\pi^2}  
\int_{V(S)} d^3 \eta^i \fr{(\vec x - \vec \eta)^i}{|\vec x - \vec \eta|^4}
$$
\be
=  {\rm i} \fr{12 ab}{4\pi^2} \Omega_3 (\vec x; S) ,
\label{45a}
\ee
where $\Omega_3 (\vec x; S)$ is the hypersolid angle comprised by
the point $\vec x$ and the surface $S$ associated to the membrane.
This is defined by $dV = R^3 d\Omega_3$, where $dV$ is the element of the volume
enclosed by the surface $S$.

Since the above commutator is a c-number, using the Baker-Hausdorf formula,
we readily find
\be
\mu (\vec x;t) \sigma (S;t) =  
\exp \lef \{ {\rm i} \fr{12 ab}{4\pi^2} \Omega_3 (\vec x; S) \ri \} 
\sigma (S;t) \mu (\vec x;t)
\label{46}
\ee
This is the dual algebra satisfied by the topological excitation and membrane
creation operators in D=5 BF theory, that generalizes the corresponding
expressions in D=2,3 and 4, given respectively by (\ref{30}), (\ref{31}) and
(\ref{32}).

We now construct the creation operator for the composite state carrying
both topological charge and tensor magnetic flux along a closed membrane 
along the surface $S$. 
We choose $S_x$ to be a 
sphere of radius $R$ centered at $\vec x$ 
and place the topological charge in the center, namely, 
\be
\psi(x;S_x;t) = \lim_{\vec x \rightarrow \vec y}
\mu(\vec x, t) \sigma(S_y, t)
\label{47}
\ee
Using the fact
that $\Omega_3 (\vec x; S_y) - \Omega_3 (\vec y; S_x) = 4 \pi^2 
\epsilon \lef (\Omega_3 (\vec x; S_y) \ri ) $, where $\epsilon(x)$ is the 
sign function, we obtain from (\ref{46})
\be
\psi(x;S_x;t) \psi(y;S_y;t) = e^{i\ 12 ab \epsilon 
\lef (\Omega_3 (\vec x; C_y) \ri )}
\psi(y;S_y;t) \psi(x;S_x;t) .
\label{48}
\ee
This relation shows that a topologically charged membrane, created by
the operator $\psi$, defined in (\ref{47}), possesses generalized statistics
determined by the parameters $a$ and $b$. For a suitable choice of the 
parameters $a$ and $b$, we can obtain 
fermionic states even though we are in the framework of a purely bosonic
BF theory.

\section{Bosonic Membranes and Strings Coupled to the Generalized
Chern-Simons Theory in D=5}
\setcounter{equation}{0}

\subsection{The Topological Charge Operator}

Let us consider in this section the generalized Chern-Simons theory coupled to 
a membrane and an external source $j^\mu$, described by the lagrangian given by
(\ref{15}). We shall firstly study the topological excitations creation operator.
Inspired in (\ref{24}), we write
\be
\mu(\vec x,t) = \exp \lef \{-i \fr{a}{3\!} \int_{-\infty,L}^{\vec x} d \xi^i 
A^{i} (\vec \xi, t) \ri \}, 
\label{49}
\ee
Let us evaluate the commutation relation of this operator with the topological charge
$Q$, which in the present theory is also given by (\ref{36}). For this, we need the 
canonical commutation rules for the D=5 Chern-Simons theory, namely
$$
[B^{ijk}(\vec x, t), A^{l}(\vec y, t)] = {\rm i} \epsilon^{ijkl} 
\delta^4(\vec x -\vec y) 
$$
\be
[B^{ijk}(\vec x, t), B^{lmn}(\vec y, t)] = [A^{i}(\vec x, t), A^{j}(\vec y, t)] = 0 ,
\label{50}
\ee
which, again, have been obtained by eliminating the second class constraints 
without fixing any gauge condition.
Calling $\mu(\vec x,t) \equiv e^{\beta (\vec x,t)}$ and using (\ref{50}), we find
\be
[Q, \beta (\vec x,t)] = a \int d^4y \int_{-\infty}^{\vec x} d \xi^i \del^i_{(\xi)}
\delta^4(\vec \xi -\vec y) = a
\label{51}
\ee
Since this is a c-number, we immediately conclude that
\be
[Q,\mu] = a \mu ,
\label{52}
\ee
implying that the operator given by (\ref{49}) indeed creates topologically charged
excitations, namely eigenstates of $Q$.

A very interesting special case of the theory described by (\ref{15}) is the one when 
the source $j^\mu$ is given by
\be
j^\mu=\oint_C d\xi^\mu\delta^5(x - \xi) ,
\label{53}
\ee
where $C$ is the border of the universe-sheet of a string. For a
closed spatial string, we have $j^0 = 0$, implying,
according to  (\ref{16}), that this type of string
does not bear any topological charge. For an
open spatial string, however, $j^0 \neq 0$ and we see that it carries
a nonvanishing topological charge. Notice that, in this case,
$C$ is still closed despite of the fact that the string is open.
We conclude that, in this situation, the operator $\mu(\vec x,t)$ in (\ref{49})
creates an open string with extremity at the point $\vec x$, along the line $L$.

\subsection{The Membrane Creation Operator}

Let us consider now the operator creating the quantum field configuration associated to
the membrane coupled to the generalized Chern-Simons theory in D=5. From (\ref{18})
and (\ref{19}), we see that there is a 2-tensor magnetic field attached to the 
membrane coupled to the tensor field in (\ref{15}). 
Following (\ref{29}) we write the membrane creation operator
in the present case as $\sigma(S;t) \equiv e^{\alpha (S;t)}$ where $\sigma (S;t)$ 
is given by expressions identical to (\ref{38}) and (\ref{39}),
which we had in D=5 BF theory. Using (\ref{3}) and the canonical commutation rules
(\ref{50}), we find that expressions (\ref{41}), (\ref{42}),(\ref{43}) and (\ref{44})
also hold here. These imply that also here the operator $\sigma (S;t)$ creates
the correct eigenstates of the magnetic field attached to the quantum membrane.

\subsection{Duality, Generalized Statistics and Fermionization}

Using the canonical commutation rules (\ref{50}), and following the same steps as
in the previous section, it is easy to see that the  topological excitation
operator $\mu(\vec x,t)$, given by (\ref{49}) and the membrane operator $\sigma (S;t)$,
given by (\ref{39}) satisfy precisely the same relations expressed in (\ref{45})
to (\ref{48}). In particular the dual algebra (\ref{46}) is the same. As a consequece,
also the composite states of a membrane and a point topological charge have
generalized statistics that may be, in particular, fermionic. Of course the previous 
remarks also hold when $\mu$ may be regarded as the creation operator of an open 
string.

\section{Perspectives}

The general duality structure investigated here has proved to be completely 
general and can be extended for any spacetime dimension. It may also be applied 
to extended objects of any dimensionality. The formalism introduced here
shall certainly be required whenever a quantum description of extended objects
such as strings or membranes coupled to a field theory should be needed.
The full quantization of an extended object (p-brane) is in general plagued by
difficulties, usually associated with anomalies. These, in a light-cone gauge 
formulation, would appear in the algebra of generators of the Lorentz group.
Only superstrings and the supermembrane in D=11 are known to be free of these
anomalies \cite{ps}. Nonsupersymmetric p-branes, on the other hand, are known to
suffer from conformal anomalies in general. The absence of supersymmetry anomalies 
is associated to the presence of massless states in the spectrum \cite{massz}. It is
precisely in connection to this fact that the present formulation is potentially useful
for the full quantization of p-branes. Indeed, dual order-disorder algebras such as 
(\ref{46}), (\ref{30}), (\ref{31}) and (\ref{32}) have been shown \cite{km} to be
closely related to a criterion for the presence of massless states in the spectrum.
This actually happens whenever the expectation values of both operators satisfying
the dual algebra vanish.

The natural extension of this work involves the evaluation of correlation functions 
of membrane and topological excitation creation operators in BF and Chern-Simons
theories, both in D=5 and D=4. This study will lead, among other things,
to a detailed description of the quantum spectrum of excitations for these extended
objects as well as of their scattering properties. It shall also certainly lead to 
very interesting unfoldings regarding bosonization in higher dimensions. 

\vfill
\eject

\section{Appendix}
\setcounter{equation}{0}

In a spacetime of dimension D=4, the spatial laplacian obeys the equation
\be
- \nabla^2 \lef [   \fr{1}{4 \pi |\vec \xi - \vec x|} \ri ] = 
\delta^3 (\vec \xi - \vec x)
\label{54}
\ee
and the above Green's function satisfies the identity
\be
\partial_{(\xi)}^i\left[\frac{1}{4\pi\vert\vec\xi-\vec x\vert}\right]
=\left\{\begin{array}{cc}
\epsilon^{ijk}\partial_{(\xi)}^j\varphi^k(\vec\xi-\vec x)
;&\vec\xi\not\in V_L\\
\int_{-\infty,L}^{\vec x}d\eta^i\delta^3(\vec\eta-\vec\xi)
;&\vec\xi\in V_L\\
\end{array}\right .  ,
\label{55}
\ee
where $V_L$ is a cone of infinitesimal angle with vertex at $\vec \xi
= \vec x$ and axis along the line $L:(-\infty,\vec x)$ and $\vec
\varphi = \frac{1-\cos\theta}{r\sin\theta}\hat\varphi$, with $r=
\vert\vec\xi-\vec x\vert$. 

In a spacetime of dimension D=5, conversely, we have for the spatial laplacian 
\be
- \nabla^2 \lef [   \fr{1}{4 \pi^2 |\vec \xi - \vec x|^2} \ri ] = 
\delta^4 (\vec \xi - \vec x)
\label{56}
\ee
Inside a hypercone containing $L$ and with the tip in $\vec x$ we have,
accordingly, the identity
\be
\partial_{(\xi)}^i\left[\frac{1}{4\pi^2 \vert\vec\xi-\vec x\vert^2}\right] =
\int_{-\infty,L}^{\vec x}d\eta^i\delta^4(\vec\eta-\vec\xi),
\label{57}
\ee
which can be easily verified by applying $\partial_{(\xi)}^i$.
Observe now that, because of the $\delta$-function,
(\ref{45}) is non vanishing only inside $V_L$, hence we can use
(\ref{57}), in order to establish (\ref{45a}).

\section{Acknowledgements}

This work has been supported in part by CNPq, FAPERJ 
and \\ PRONEX - 66.2002/1998-9.

\end{document}